\documentclass[prl,twocolumn,floatfix,superscriptaddress,amsmath,citeautoscript,aps,longbibliography]{revtex4-2}
\pdfoutput=1
\usepackage[T1]{fontenc}\usepackage[latin1]{inputenc}
\usepackage{dcolumn,bm,graphicx,color,booktabs,microtype,afterpage} \graphicspath{{Figures/}}
\usepackage[charter,greekuppercase=italicized]{mathdesign}
\renewcommand{\figurename}{Fig.}
\renewcommand{\tablename}{Table}
\makeatletter\renewcommand{\fnum@figure}[1]{\figurename~\thefigure.}\makeatother
\makeatletter\renewcommand{\fnum@table}[1]{\tablename~\thetable.}\makeatother
\newcount\hh \newcount\mm
\hh=\time \divide\hh by 60
\mm=\hh \multiply\mm by 60 \mm=-\mm
\advance\mm by \time
\def\now{\number\hh:\ifnum\mm<10{}0\fi\number\mm}
\usepackage[colorlinks,plainpages=false,linkcolor=blue,urlcolor=blue,citecolor=blue,pdfpagemode=UseNone,pdfstartview=FitBH]{hyperref}
\newcommand{\magenta}[1]{\textcolor{black}{#1}}
\newcommand{\green}[1]{\textcolor{black}{#1}}

\begin{document}

\title{Phonon topology and winding of spectral weight in graphite}

\author{N.~D.~Andriushin}
\affiliation{Institut f{\"u}r Festk{\"o}rper- und Materialphysik, Technische Universit{\"a}t Dresden, D-01069 Dresden, Germany}
\author{A.~S.~Sukhanov}
\affiliation{Institut f{\"u}r Festk{\"o}rper- und Materialphysik, Technische Universit{\"a}t Dresden, D-01069 Dresden, Germany}
\author{A.~N.~Korshunov}
\affiliation{European Synchrotron Radiation Facility (ESRF), BP 220, F-38043 Grenoble Cedex, France}
\author{M.~S.~Pavlovskii}
\affiliation{Kirensky Institute of Physics, Siberian Branch, Russian Academy of Sciences, Krasnoyarsk 660036, Russian Federation}
\author{M. C. Rahn}
\affiliation{Institut f{\"u}r Festk{\"o}rper- und Materialphysik, Technische Universit{\"a}t Dresden, D-01069 Dresden, Germany}
\author{S.~E.~Nikitin} 
\thanks{Corresponding author: stanislav.nikitin@psi.ch}
\affiliation{Quantum Criticality and Dynamics Group, Paul Scherrer Institut, CH-5232 Villigen-PSI, Switzerland} 
\affiliation{Laboratory for Neutron Scattering and Imaging, Paul Scherrer Institut, CH-5232 Villigen-PSI, Switzerland}

\begin{abstract}

The topology of electronic and phonon band structures of graphene is well studied and known to exhibit a Dirac cone at the K point of the Brillouin zone. Here, we applied inelastic x-ray scattering (IXS) along with \textit{ab initio} calculations to investigate phonon topology in graphite, the 3D analogue of graphene. We identified a pair of modes that form a very weakly gapped linear anticrossing at the K point that can be essentially viewed as a Dirac cone approximant. The IXS intensity in the vicinity of the quasi-Dirac point reveals a harmonic modulation of the phonon spectral weight above and below the Dirac energy, which was previously proposed as an experimental fingerprint of the nontrivial topology. We illustrate how the topological winding of IXS intensity can be understood in terms of atomic displacements, and highlight that the intensity winding is not in fact sensitive in telling quasi- and true Dirac points apart.

\end{abstract}

\maketitle

\textit{Introduction.}
The mathematical concept of topology has become a key concept of modern condensed matter physics~\cite{bhattacharjee2017topology, wang2017topological}. Topology allows one to classify continuous manifolds and thus provides a new organizing principle for the electronic band structure of solids~\cite{bansil2016colloquium, narang2021topology}. The textbook example is the honeycomb structure of graphene, whose triangular Bravais lattice precludes a hybridization of the orthogonal electronic states originating from its two-atomic basis. Electrons in the vicinity of the resulting linear band-crossing behave as massless quasiparticles described by the Dirac equation and are responsible for many unconventional transport effects.

Notably, the concept of topology does not depend on details of quantum statistics and is just as applicable to bosonic quasiparticles like as phonons~\cite{liu2020topological} or magnons~\cite{mcclarty2022topological}. The theoretical study of phonon topology has recently attracted special interest. Significant advances include the prediction of a surface arc-state in hexagonal WC-type materials~\cite{xie2019phononic}, quadratic nodal lines and hybrid nodal rings in AgZr~\cite{zhou2021hybrid} and topological gimbal phonons in $T$-carbon~\cite{you2021topological}. Experimentally, the topology of lattice fluctuations is also being actively studied in artificial mesoscopic structures with linear dimensions of the order of mm, resulting in resonance frequencies in the kHz range. These developments have in fact coined a new research field -- topological acoustics~\cite{yang2015topological, zhang2018topological, xue2022topological}. However, in conventional crystals with interatomic distances of several~\AA~and vibration frequencies of $\approx$~10~THz, the unambiguous identification of topological phonon crossings remains a challenge, and only few reports have been published so far~\cite{miao2018observation, nguyen2020topological, jin2022chern}. A common approach in this field is to characterize the dispersion in the vicinity of crossing points, and compare this data with \textit{ab-initio} calculations. Non-trivial topological properties can then be inferred from the analysis of the calculated phonon dispersions.  Crucially, this does not provide an immediate experimental probe of topology, rather than a verification of the density functional model.

Recently, Jin~\textit{et al.}~\cite{jin2022chern} proposed an experimental method to probe topological character of a phonon crossing by measuring the spectral weight in its vicinity. The same approach is already widely used in application to the magnon band structure~\cite{shivam2017neutron, elliot2021order, scheie22, nikitin2022thermal}. It reads that on closed momentum-space contours around the crossing point, the intensity of phonon excitations is modulated, with a number of minima and maxima related to the Berry phase of the point. This phenomenon results from the modulation of phonon eigenvectors around the crossing point. It was therefore proposed as a direct experimental probe of the phonon topology.

\begin{figure*}[tb]
\center{\includegraphics[width=1\linewidth]{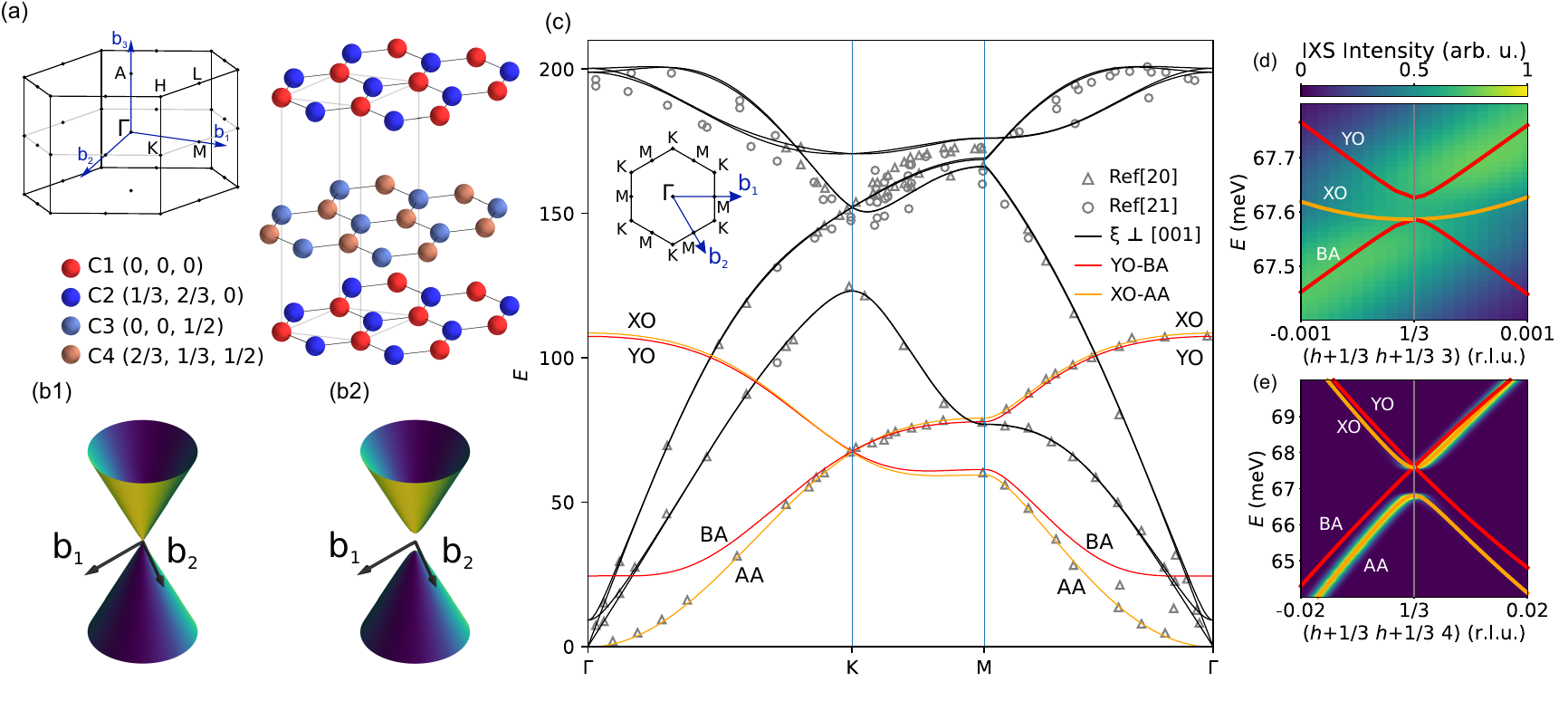}}\vspace{-12pt}
  \caption{~Crystal structure and phonon excitations in graphite. 
   (a) The crystal structure, formed from an AB stacking of honeycomb layers, and the first Brillouin zone. 
  (b1), (b2) Schematic representation of the phonon dispersion around the K point, showing the gapless Dirac cone in graphene (b1), which gaps out due to interlayer coupling in graphite (b2).
  (c) Comparison of our DFT calculation (solid lines) of the phonon dispersion in graphite with previous IXS measurements~\cite{mohr2007phonon, Maultzsch_2004}. The first BZ of graphite  within the $(hk0)$ plane is shown in the inset. 
  (d), (e) Detailed view of the YO--BA (red line) and XO-AA (orange line) modes in the vicinity of K along $(hh3)$ and $(hh4)$ paths of reciprocal space (see details of reciprocal-space trajectories in Sec.~S2 of SM~\cite{SI}). The color map indicates the simulated IXS intensity. 
  }
  \label{Fig1} \vspace{-12pt}
\end{figure*}

In the present Letter, we apply this method to study low-energy phonons in the vicinity of the $K$ point in graphite using inelastic x-ray scattering (IXS) and density-functional theory (DFT) calculations [see Sec.~S1 of Supplemental Material (SM)~\cite{SI} for method details] and demonstrate the limitations of such approach, which were not previously pointed out. Graphite consists of graphene layers stacked with a relative shift of [2/3, 1/3, 0] along the $ab$-plane [known as the AB stacking, Fig.~\ref{Fig1}(a)]. Although its phonon dispersions were previously mapped in detail throughout the entire Brillouin zone (BZ)~\cite{Maultzsch_2004, mohr2007phonon}, the distribution of the spectral weight, which carries the crucial information on the eigenvector modulation, has not been addressed. We identified both theoretically and experimentally that there exists a pair of low-energy modes that form a weakly gapped pseudo-Dirac cone at the K point. Our IXS data demonstrate the antiphase distribution of the phonon spectral weight above and below the pseudo-Dirac point, in perfect agreement with the theoretical predictions applied for a true Dirac point. Our DFT calculations illustrate that the microscopic origin of this winding pattern lies in the relative phase shift of out-of-plane oscillations of two atoms within the honeycomb layers. The latter remains intact upon opening of a barely noticeable gap at the Dirac point caused by small but finite perturbative interlayer interaction.

\textit{Overview of the phonon dispersion}.
We start the presentation of our results with a brief overview of the phonon dispersion of graphite, shown in Fig.~\ref{Fig1}(c)~\cite{mohr2007phonon, Maultzsch_2004, bosak2007elasticity, gruneis2009phonon}. Our DFT calculations of the dispersion fully corroborate previous measurements and calculations~\cite{michel2008theory, li2020topological}. Based on this excellent agreement, we can use this model to simulate the distribution of the phonon spectral weight and relate it to the structural fluctuation patterns in the vicinity of the K point.

\begin{figure*}[tb]
\center{\includegraphics[width=1\linewidth]{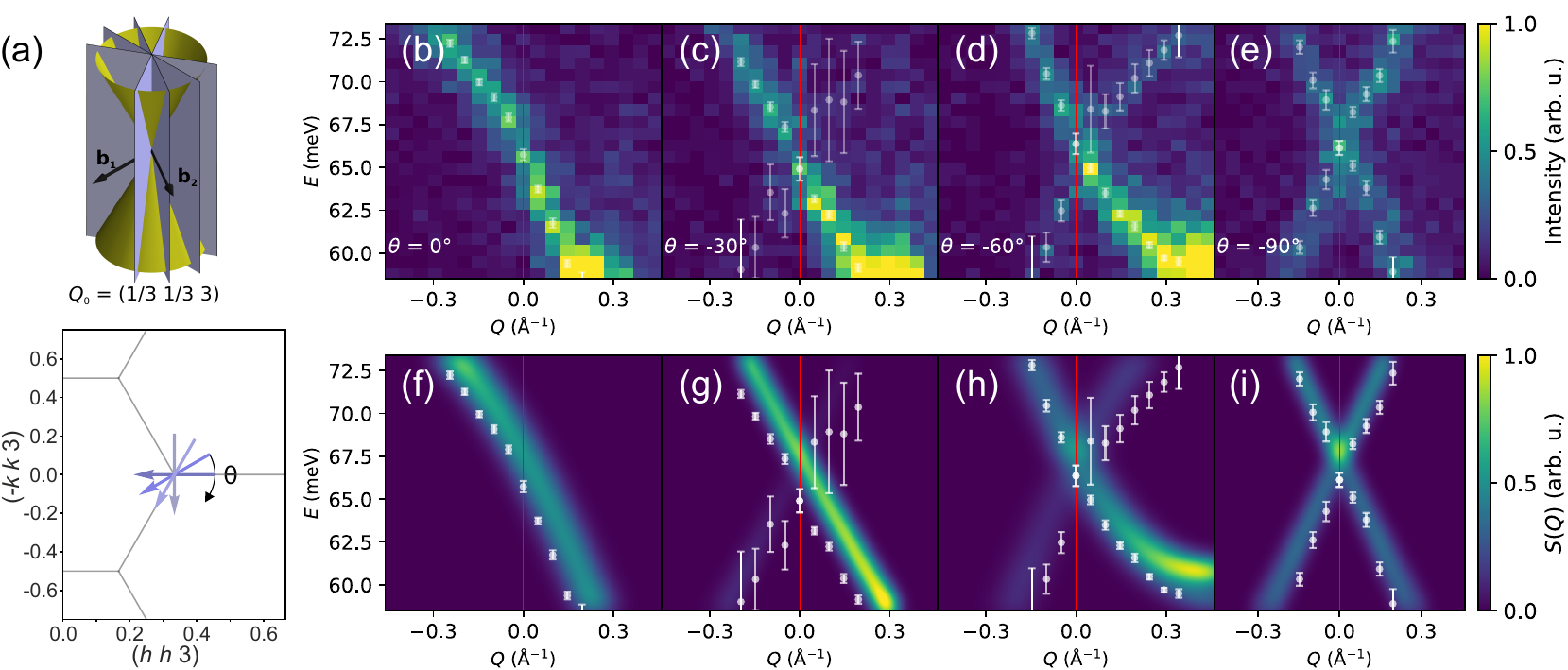}}\vspace{-12pt}
  \caption{~Phonon dispersion in the vicinity of the K point at $\mathbf{Q} = (1/3~1/3~3)$. (a) Schematic of the slices through 4D energy-momentum reciprocal space. (b)--(e) IXS data along selected radial directions centered at the K point. Note that additional intensity at $E < 61$~meV in (b) and (c) is due to contribution from the second crystal domain in the sample. (f)--(i) Corresponding simulated IXS spectra. \magenta{White symbols denote the fitted peak positions with 95\% confidence interval~\cite{SI}.}
  }
  \label{Fig2}\vspace{-12pt}
\end{figure*}

We can broadly categorize all modes in Fig.~\ref{Fig1}(c) into two groups: low-energy $c$-polarized modes (black lines) and high-energy modes with $\bm{\xi}\perp{}$[001] (orange and red lines). Visual inspection of Fig.~\ref{Fig1}(c) suggests that the dispersion has several crossings at the K point at $E~=~$67, 152 and 170~meV. However, the modes close to K points at energies $\approx\!152$ and 170~meV have shallow dispersion and remain linear only within a narrow energy range of $\lesssim$1.5~meV, which means they can be classified as not ``clean'' crossings according to criteria defined in~\cite{li2021computation}. 
Therefore, we focus on the low-energy part of the spectrum. It contains four modes: the acoustic (AA) and the low-energy pseudo-acoustic (BA) modes, and two almost degenerate downward-dispersing optical branches, (YO) and (XO), at 65--110~meV. The AA and BA modes are related to in- and anti-phase fluctuations of carbon layers along the $c$ axis, and their splitting is determined by the strength of the interlayer coupling. Furthermore, the structure factor $F^{(s)} (\mathbf{Q})$ has a profound impact on the observable intensity, which allows us to separate the modes: (i) both modes have finite intensities only at non-zero $l$ ($l$ is the reciprocal space coordinate); (ii) due to the structure factor modulation, AA and BA intensities are harmonically modulated along the $l$ direction and acquire maximal intensity at even and odd $l$, respectively. 

The acoustic and optical $c$-polarized phonon modes in graphene are known to form a Dirac cone centred at the K point as shown in Fig.~\ref{Fig1}(b1)~\cite{li2020topological}. However, in graphite, a finite interlayer coupling breaks the inversion symmetry between the carbon atoms within a honeycomb layer and gaps out the otherwise degenerate states at the K point, causing the typical anticrossing behaviour [Fig.~\ref{Fig1}(b2)]~\cite{michel2008theory}. To visualize this effect, in Fig.~\ref{Fig1}(e) we show the calculated IXS intensity for the slice along the $(hh4)$ direction that maximizes intensity of XO and AA modes. Here, a gap of $\sim$0.8~meV is clearly visible at the K point. 
The YO and BA modes are best visible at odd $l$ and almost degenerate at the K point with an extremely small gap of $\sim$50~$\mu$eV [Fig.~\ref{Fig1}(d)]. We should stress out that these modes do not form a true topologically protected crossing as the mode repulsion is induced by a weak interlayer interaction. However, the splitting of the YO--BA mode pair is in essence irrelevant for the given slope of the linear dispersion, i.e. everywhere except for the vanishingly narrow momenta range of $\kappa_0 \approx \pm 10^{-3}$ r.l.u. as we discuss in details in Sec. S4 of SM~\cite{SI}.
Below we demonstrate that the distribution of the phonon spectral weight in the vicinity of the K point but at any momenta greater than $\kappa_0$ follows the characteristic intensity winding behavior, which was deemed to only accompany a true Dirac point. We then focus on the K point of $\mathbf{Q}=(1/3,1/3,3)$, where the structure factor maximizes the scattering intensity of the weakly repulsed YO and BA modes.

\begin{figure}[tb]
\center{\includegraphics[width=1\linewidth]{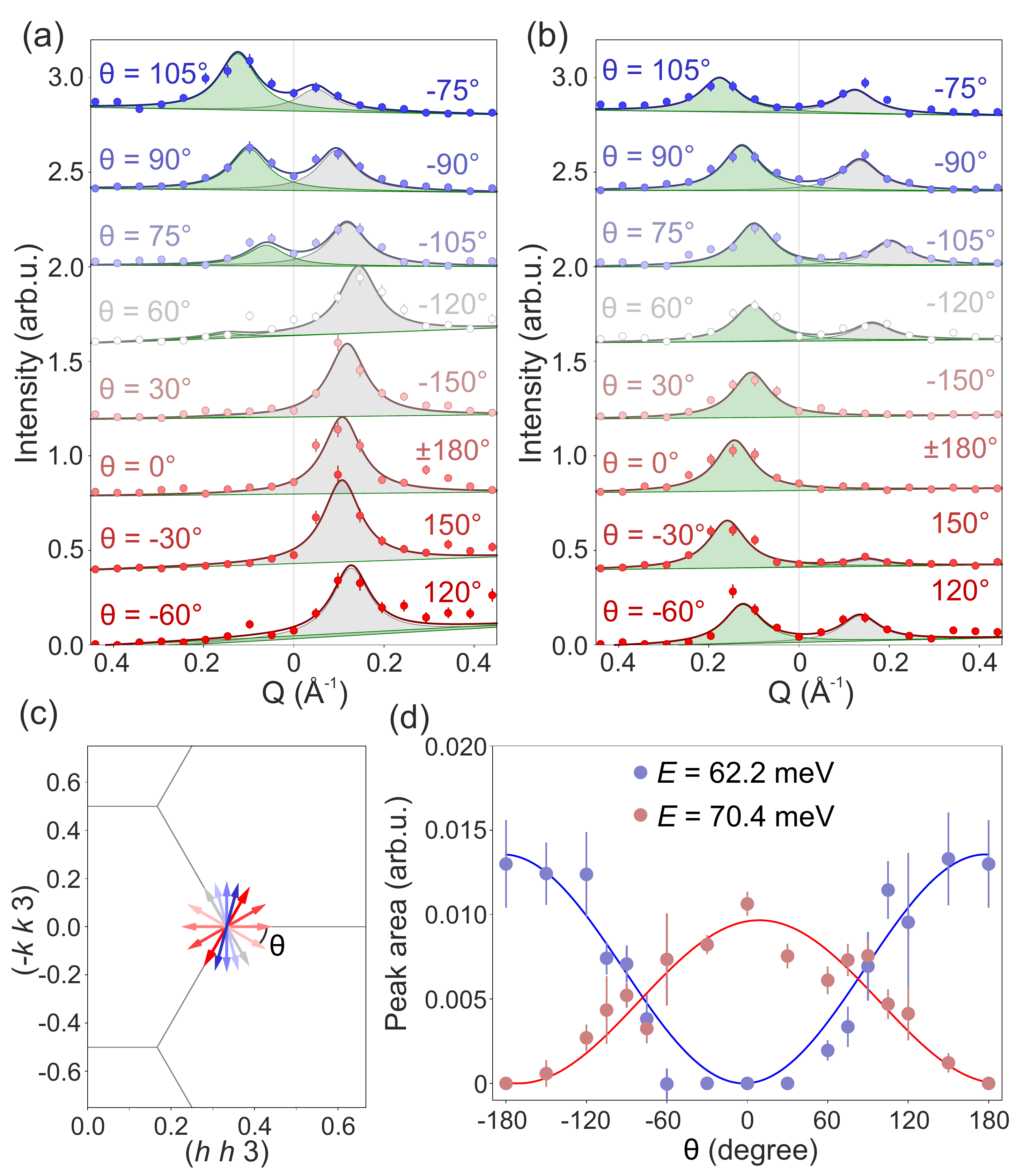}}\vspace{-12pt}
  \caption{~Modulation of the phonon spectral weight in graphite. 
  \magenta{(a), (b) Radial momentum profiles of IXS intensity through the K point in the $(hk3)$ plane at $E = 62.2$~(a) and 70.4~meV~(b). Solid lines represent a complete fit to the data that includes phonon contributions (Lorentzian peaks) and a linear background. Green and gray shaded areas show the phonon peaks measured on the opposite trajectories off the K point.
  The data are offset by +0.4 units for visual clarity. 
  (c) Trajectory of the momentum profiles shown in (a) and (b), and the definition of the angle $\theta$. 
  (d) $\theta$-dependence of the peak area extracted from the IXS measurements. Solid lines are fits with Eq.~\eqref{eq_cos}.}
  }
  \label{Fig3}\vspace{-12pt}
\end{figure}

\textit{IXS spectra.}
Figure~\ref{Fig2} illustrates the variation of the IXS spectra around $\mathbf{Q}=(1/3,1/3,3)$. The orientation of four energy-momentum slices with respect to the phonon Dirac cone is visualized in Fig.~\ref{Fig2}(a). Figure~\ref{Fig2}(b) shows the longitudinal slice along the $(hh3)$ direction. It features a single, linearly dispersing phonon mode that crosses the K point at Dirac energy $E_{\rm Dir} = 65.6(2)$~meV, slightly below the calculated value of 67\,meV. Figures~\ref{Fig2}(c) and \ref{Fig2}(d) show spectra along two radial paths, rotated by 30$^{\circ}$ and 60$^{\circ}$ to the $(hh3)$ direction, respectively. The signal in Fig.~\ref{Fig2}(c) looks rather similar to the $(hh3)$ path, with only a faint trace of the second mode, which becomes clearly visible in the data in Fig.~\ref{Fig2}(d).  Figure~\ref{Fig2}(e) shows the radial slice along the $(1/3-k,~1/3+k,~3)$ direction, i.e. orthogonal to $(hh3)$. It exhibits an X-shaped crossing at the K point. The two crossing modes have similar IXS intensity over the whole energy range, in agreement with the simulated spectrum, Fig.~\ref{Fig2}(i). \green{Taken together, our data indicate good agreement with the simulated spectra and show that YO and BA modes touch at the K point within our instrumental resolution, and exhibit the linear dispersion within $\approx\pm 0.1$~\AA$^{-1}$, followed by some extended quasi-linear regime up to $\approx\pm 0.3$~\AA$^{-1}$ (see \cite{SI} for a strict criteria of nonlinearity).} The spectral weight is gradually redistributed over both modes under rotation around the K point. All DFT results are in good agreement with these observations. We further comment that the calculated gap at K point for YO--BA crossing is only $\sim$50~$\mu$eV, which is well below the experimental resolution and therefore cannot be resolved in our data.

Having discussed the dispersion around the K point, we turn to the quantitative analysis by plotting the IXS intensity on radial constant-energy paths through the K point. Figures~\ref{Fig3}(a,b) show a series of these slices at 62.2 and 70.4\,meV, below and above the Dirac energy $E_{\rm Dirac}\pm\ 4$~meV. Panel~\ref{Fig3}(c) illustrates the trajectories of these scans and our definition of an angle $\theta$ with respect to the $(hh3)$ direction. Each slice was fitted using two Lorentzians to quantify the spectral weight of each phonon mode. In Fig.~\ref{Fig3}(d), we summarize these peak areas as a function of the angle $\theta$. This reveals harmonic oscillations described by 
\begin{align}
    I(\theta) =\ a_0[1 \pm\ \mathrm{cos}(\theta)],
    \label{eq_cos}
\end{align}
 for energies above and below the Dirac energy, respectively, with $a_0$ as a constant prefactor. This result is perfectly consistent with our DFT simulations and the theoretically predicted modulation for a Dirac point~\cite{jin2022chern}. Notably, a similar modulation of spectral weight was observed in several magnonic honeycomb ferromagnets including CoTiO$_3$~\cite{Yuan_2020, elliot2021order}, elemental Gd~\cite{scheie22} and CrBr$_3$~\cite{nikitin2022thermal}.

\begin{figure}[t]
\center{\includegraphics[width=1\linewidth]{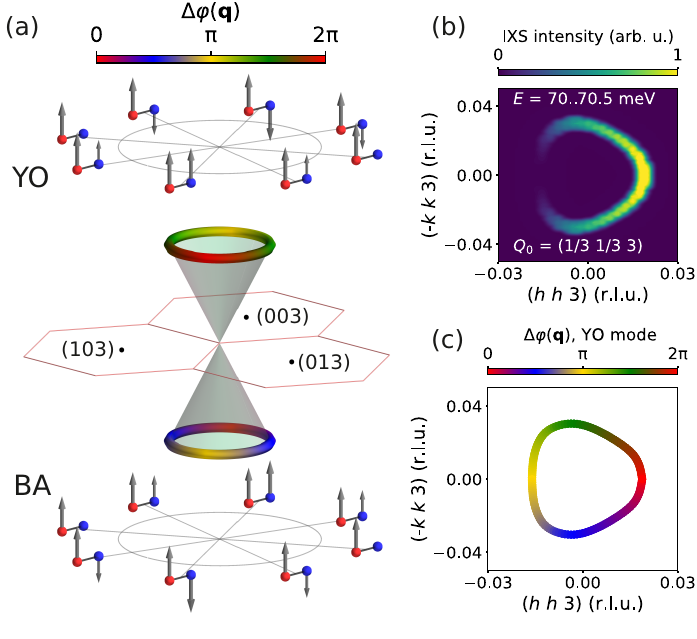}}
  \caption{~Microscopic origin of the intensity winding.
    (a)~Schematic representation of the pseudo-Dirac cone. Colorplots at the cone bases represent phase difference of two atomic displacements, $\Delta\varphi(\mathbf{Q})$. Two sketches above and below the pseudo-Dirac cone demonstrate the calculated atomic displacements of two atoms within the honeycomb layer, $\xi_1$ and $\xi_2$ at \magenta{several equidistant $\mathbf{Q}$ points.} 
    (b)~IXS intensity distribution within the $(hk3)$ plane calculated at $E = [70, 70.5]$~meV, above the pseudo-Dirac point.
    (c)~Phase difference of two atomic displacements, $\Delta\varphi(\mathbf{Q})$ calculated for the same energy. 
  }
  \label{Fig4}
  \vspace{-12pt}
\end{figure}

\textit{Microscopic origin of the intensity modulation.}
% Our IXS data shown in Fig.~\ref{Fig3}(d) demonstrate a smooth modulation of the phonon spectral weight. To provide an intuitive picture of how this winding relates to atomic displacements, we further examine the results of our DFT simulations. The dispersions of the YOBA and ZO$_2$ modes were calculated on a small circle around the $K$ point (diameter $|\Delta{}Q| = 0.02$~\AA$^{-1}$), just above and below the crossing. Figure~\ref{Fig4}(a) shows that the calculated dispersion remains almost flat within the considered $\mathbf{Q}$ range. In accordance with naive expectations {\color{red}(see Sashas comment in Overleaf)}, both modes are polarized along the $c$-axis (i.e. only the $z$-components of the eigenvectors are finite). As a concequence, the eigenvector of each phonon mode close to the Dirac point can be characterized by four complex numbers, which represent the displacements of four atoms (two atoms per layer) within the unit cell.
Our IXS data shown in Fig.~\ref{Fig3}(d) demonstrate a smooth modulation of the phonon spectral weight. To provide an intuitive picture of how this winding relates to atomic displacements, we further examine the results of our DFT simulations. The dispersions of the YO and BA modes were calculated on a closed contour around the K point (diameter $|\Delta{}\mathbf{Q}| = 0.02$~\AA$^{-1}$), just above and below the pseudo-crossing. Within the considered $\mathbf{Q}$ range, the dispersion of both modes remain almost flat. Moreover, because the modes are polarized along $c$ axis [Fig.~\ref{Fig1}(c)], the eigenvector of each phonon mode close to the pseudo-Dirac point (the same, in principle, applicable to a true Dirac point as well) can be characterized by four complex numbers, which represent the displacements of four atoms (two atoms per layer) within the unit cell.

However, because we are interested in the in-plane dispersion, the problem can be simplified. Given that at every $\mathbf{Q}$-point of interest the two layers exhibit antiphase vibrations, these modes can be described using only the displacements of the two atoms in a honeycomb layer, $\xi^z_1(\mathbf{q})$ and $\xi^z_2(\mathbf{q})$. An analysis of the eigenvectors shows that $|\xi^z_1(\mathbf{q})| \approx\ |\xi^z_2(\mathbf{q})| = \xi^z_0$ and thus, the phase difference between the eigenvectors of the two atoms, $\Delta\varphi(\mathbf{q})\ = \mathrm{arg}[\xi^z_1(\mathbf{q})] - \mathrm{arg}[\xi^z_2(\mathbf{q})]$, is the primary parameter that controls the distribution of the spectral weight around the K point. To demonstrate this, in Fig.~\ref{Fig4}(a) we plot the calculated $\Delta\varphi(\theta)$ for energies above and below the Dirac energy. This shows a continuous rotation around the K point, with a relative shift of $\pi$ for energies above and below the Dirac point. To visualize how the phase difference relates to atomic motion, in Fig.~\ref{Fig4}(a) we present evolution of the atomic displacements, $\xi^z_1(\mathbf{q})$ and $\xi^z_2(\mathbf{q})$, for several $\mathbf{Q}$ point on a circle around K for both energies.

Now we show analytically, how the phase difference causes the intensity redistribution. We substitute \magenta{the eigenvectors in a form} $\xi^z_i(\mathbf{q}) = |\xi^z_0| e^{i\varphi_i(\mathbf{q})}$ in the standard equation for the IXS intensity~\cite{SI} and find that $|F^s(\mathbf{q})|^2 \propto\ 1+\mathrm{cos}(\Delta\varphi(\mathbf{q}))$. Since $\Delta\varphi~\propto~\theta$ [see Fig.~\ref{Fig4}(c)] the IXS intensity exhibits the gradual harmonic modulation [Fig.~\ref{Fig4}(b)]. That provides an intuitive microscopic explanation for the observed cosine modulation of the intensity around the Dirac point in graphite, in accordance with Eq.~\eqref{eq_cos} \magenta{(see Sec.~S3 of SM~\cite{SI} for the detailed derivation).}

\textit{Discussion and Conclusion.}
Topological bosonic quasiparticles have been actively studied in condensed matter physics over the last years~\cite{Vishwanath_2013, li2020topological, Yuan_2020,li2021computation, McClarty_2022}. The majority of experimental studies were aimed to determine the \textit{dispersion} in the vicinity of a crossing point of interest, while the intensity distribution is rarely inspected in detail~\cite{miao2018observation, nguyen2020topological,zhang2020magnonic, chen2018topological}. However, one striking feature of topological crossings is that some physical quantity should exhibit a winding around it. Usually, the winding is associated with a pseudo-spin of a low-energy Hamiltonian that describes the dynamics of a model close to the crossing point~\cite{Xiao_2010}. 
Coming back from pseudo-spin to phonon or magnon terminology, this corresponds to a rotation (in a broad sense) of the eigenvector on a close path around the crossing point. This will be reflected in a physical observable like the spectral weight in a scattering experiment, thus providing a direct measure of topological charge~\cite{jin2022chern}.

Here, we have applied this idea to address the low-energy linear phonon crossings in graphite, one of the simplest honeycomb system. Our DFT calcultions indicate that the interlayer coupling is a relevant perturbation that breaks the symmetry between the two carbon atoms within a honeycomb layer and hybridizes the otherwise crossing YO--BA modes at the K point. The gap is controlled by strength of interlayer coupling and amounts to only $\sim$50~$\mu$eV. Because the interlayer coupling in graphite is induced by weak van der Waals forces, it can be noticeably enhanced by hydrostatic pressure. This way one can effectively push the system towards 3D regime and increase the gap at the $K$ point. In this case, a large gap breaks the widning of the spectral weight and pseudo-Dirac states no longer exist~\cite{SI}.

%We demonstrate this effect by DFT calculations and discuss its consequences on intensity winding in Sec.~S6 of SM~\cite{SI}.

However, the resulting avoided crossing of the YO--BA modes plays a role only in the momenta scale as small as $\sim 5 \cdot 10^{-4}$ \AA$^{-1}$ away from the $K$ point in absence of external pressure. Outside this narrow region, the modes in essence behave indistinguishably from modes forming a true Dirac point, hence referred to as the pseudo-Dirac cone. Because the gap is too small to be resolved in an experiment, our IXS data show that the YO and BA modes exhibit a linear crossing at the K point, and also provide a clear evidence for \textit{intensity winding} with opposite phases above and below $E_{\rm Dirac}$. This result is in perfect agreement with the theoretical expectations for the $\pm \pi$ Berry phase on a contour surrounding a Dirac point. By analysing the displacement vectors, we show that the relevant physical quantity which winds and produces this peculiar intensity distribution is $\Delta\varphi$, the phase difference between the oscillation of the two atoms in the honeycomb plane.

Therefore, the approach of identifying the intensity winding cannot be solely applied when the topological properties of the band structure are addressed. Our results indicate that the experimental observation of intensity modulation proposed in~\cite{jin2022chern} for the phonon bands and already widely used in application to the magnon bands~\cite{nikitin2022thermal, elliot2021order, scheie22} can only serve as a fingerprint of the winding of quasiparticle eigenvector rather than prove existence of a topologically protected crossing, unless the measurements are taken on a scale of the smallest gap in the system, that is often not possible due to resolution broadening, especially in low-dimensional systems such as graphite with strong hierarchy of couplings. We have thus identified winding of the phonon eigenvectors due to proximity to a pseudo-Dirac point and show how IXS can be applied to study topological lattice excitations in condensed matter systems. 

\textit{Acknowledgments}
We thank R. Coldea for stimulating discussions. A. Bosak participated in the discussion and preparation of the experiment, however he cannot make part of the list of authors following the Resolution adopted by the ESRF Council on 22 March 2022.
We acknowledge financial support from the Swiss National Science Foundation, from the European Research Council under the grant Hyper Quantum Criticality (HyperQC), the German Research Foundation (DFG) through the Collaborative Research Center SFB 1143 (project \# 247310070) and from the European Union Horizon 2020 research and innovation program under Marie Sk\l{}odowska-Curie Grant No.~884104.  We acknowledge the European Synchrotron Radiation Facility (ESRF) for provision of synchrotron radiation facilities.

\bibliography{bibliography}

\end{document}

\onecolumngrid

\centerline{\large{\bf {Supplementary Information to accompany the article}}} 

\vskip1mm

\centerline{\large{\bf {``Phonon topology and winding of spectral weight in graphite''}}}

\vskip4mm

\centerline{N. D. Andriushin, A. S. Sukhanov, A. N. Korshunov, M. S. Pavlovskii, M. C. Rahn and S. E. Nikitin } 

\vskip20mm
\twocolumngrid

\section{Methods.}
High-resolution IXS measurements were performed at beamline ID28 (ESRF)~\cite{krisch2017inelastic}. The incident energy ($E_{\rm i} = 17.8$~keV) was selected using the Si~(9~9~9) reflection, which resulted in an effective energy resolution of $\mathrm{\Delta}E=3$~meV (full width at half maximum). A sample of Kish graphite (space group $P6_3/mmc$) was obtained commercially from \textsc{2D Semiconductors}~\cite{2d_semiconductor}. Its crystalline quality was verified by single-crystal x-ray diffraction. The measurements showed that the sample contained equally-populated crystal domains rotated by 30(2)$^{\circ}$ around the $c$ axis, with mosaicity of each domain not worse than $\approx~0.6^{^{\circ}}$. Therefore, the measured IXS spectrum is actually a superposition of the signals of either twin. Crucially, for the low-energy modes of interest, one structural domain does not contribute any phonon intensity in the vicinity of the K point of the other, meaning that our experimental data are not affected. All the measurements were conducted at room temperature.

Lattice-dynamics calculations were carried out using the projector-augmented wave (PAW) method~\cite{PhysRevB.59.1758} and density functional theory (DFT) as implemented in the \textsc{vasp} software~\cite{PhysRevB.54.11169,KRESSE199615}. The generalized gradient approximation (GGA) functional with Perdew-Burke-Ernzerhof (PBE) parametrization~\cite{PhysRevLett.77.3865} was used. A $19\times19\times6$ $k$-point mesh (Monkhorst-Pack scheme~\cite{PhysRevB.13.5188}) was used for Brillouin zone integration. The plane-wave cutoff was set to 400~eV.

The IXS scattering cross section is given by
\begin{align}
S(\mathbf{Q},E) &\propto\ \sum_{\mathbf{G, q, s}} \frac{|F^{(s)} (\mathbf{Q})|^2}{2\omega^{(s)}(\mathbf{q})}\times\delta(\mathbf{Q - q - G}) \delta(\omega - \omega^{(s)}(\mathbf{q})); \\
F^{(s)} (\mathbf{Q}) &= \sum_d \rho_{d}(|\mathbf{Q}|) \hspace{.1cm} \mathbf{Q} \cdot\ \bm{\xi}_d^{(s)} (\mathbf{q}) e^{i\mathbf{Q}\cdot\mathbf{r}_d},
\label{eq:eq1}
\end{align}
where $\omega^{(s)}$ and $\bm{\xi}_d^{(s)}$ are the energy and the polarization of the phonon with the mode index $s$, $\mathbf{Q} = \mathbf{q + G}$ is the total momentum transfer, $\mathbf{G}$ is a reciprocal lattice vector and $\mathbf{q}$ is a phonon wave-vector; $\mathbf{r}_d$ is the position vector for the atom $d$ in the unit cell, and $\rho$ is the scattering length. The phonon spectral weight was simulated using the \textsc{oclimax} software~\cite{cheng2019oclimax} which uses vibrational frequencies and polarization vectors from first-principles calculations as input. The phonon modes were obtained through construction of a supercell ($4\times4\times2$) and calculation of the force constants, as implemented in \textsc{phonopy}~\cite{TOGO20151}.

\section{Reciprocal space of graphite}

\begin{figure}[tb]
\center{\includegraphics[width=1\linewidth]{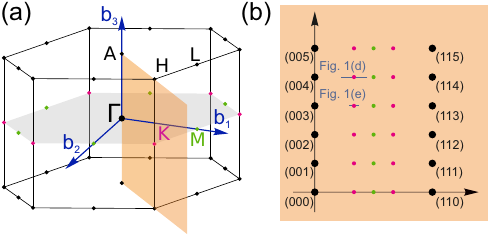}}\vspace{-12pt}
  \caption{~The scheme of reciprocal lattice of graphite. (a) The first Brillouin zone with denoted high-symmetry points. The ($hhl$) plane is highlighted by the orange rectangle, whereas the grey hexagon shows the $(hk0)$ plane. (b) The ($hhl$) plane, the blue lines correspond to ($hh$3) and ($hh$4) reciprocal paths. Note that the relative length of the reciprocal lattice vectors was arbitrary scaled for clarity and is not equal to that of graphite.
  }
  \label{SFig_reciprocal_space} \vspace{-12pt}
\end{figure}

Figure~\ref{SFig_reciprocal_space} sketches reciprocal space of graphite and visualize the directions of the energy-momentum slices that are presented in Fig.~1. of the main text. As can be seen, the reciprocal $(hhl)$ plane intersects the first BZ of a hexagonal crystal structure such that the high-symmetry points $\Gamma$, $A$, $K$, and $M$ lay in the plane. For integer $L$, the $(hh0)$ direction consequently intersects the points $\Gamma$--$K$--$M$--$\Gamma$.

\section{Derivation of the structure factor} \label{sec:eq}

As was shown in the main text, by substituting the displacement vectors in a form $\xi_d^z = \xi^z_0 e^{i\varphi_d}$ to the equation for the structure factor \ref{eq:eq1}, one can find that $|F(\bm{Q})|^2 \propto 1 + \mathrm{cos}(\Delta \varphi (\bm{Q}))$, where $\Delta \varphi (\bm{Q})$ is the oscillation phase difference of the two inequivalent carbon atoms within the honeycomb layers at a given momentum distance $\bm{Q}$ from the K point projected on the $(hk0)$ plane.

\begin{figure}[tb]
\center{\includegraphics[width=1\linewidth]{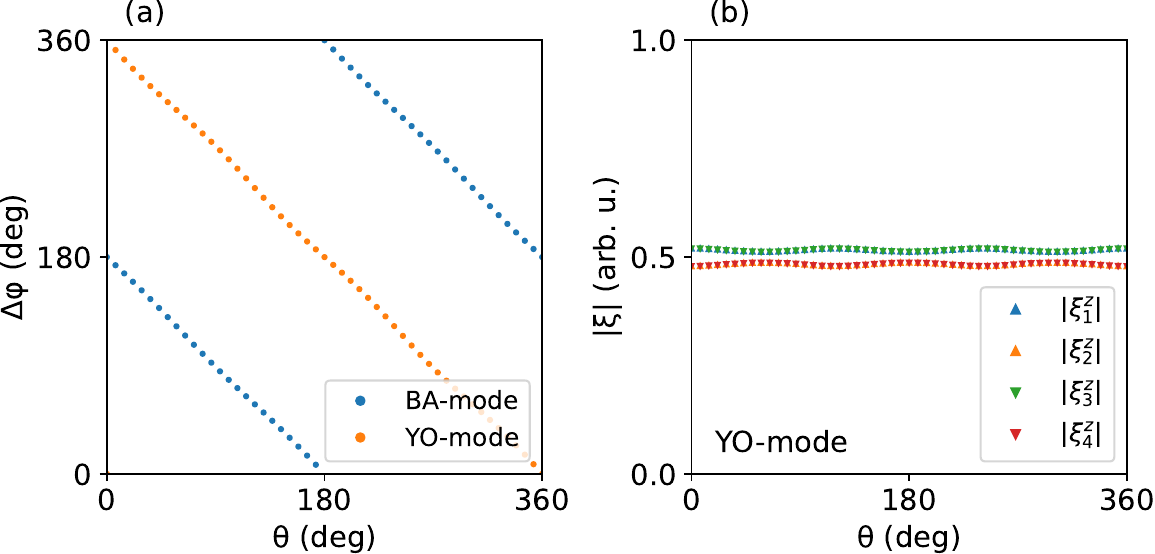}}\vspace{-12pt}
  \caption{~(a) The phase difference $\Delta\varphi$ for YO and BA modes calculated on a small circle around $\bm{Q} = (\frac{1}{3}~\frac{1}{3}~3)$ in the $(HK3)$ plane, $\Delta \mathbf{Q} = 0.01$~\AA$^{-1}$. (b) The eigenvectors amplitudes $|\xi^z|$ calculated on the same circle as in (a) for YO mode. Note that blue and orange symbols are barely visible, because their position well coincides with green and red curves.
  }
  \label{SFig_eigenvectors} %\vspace{-12pt}
\end{figure}

In this section, we provide a detailed derivation of this relation. Note, that here we are working in a close proximity to the K point, meaning that the total momentum transfer, $\bm{Q}$, can be written as a sum of two parts, $\bm{Q} = \bm{Q}_K + \delta\bm{Q}$, where the first term corresponds to the wavevector of a given K point and $\delta\bm{Q}$ is the difference between $\bm{Q}$ and $\bm{Q}_K$ with $|\delta\bm{Q}| \ll\ |\bm{Q}_K| \approx\ |\bm{Q}|$. 
%In the main text we took into account only two atoms within a honeycomb layer. Here we consider this case and also full four atomic basis of graphite to show that AB stacking does not changes the intensity modulation as long as the interlayer coupling is weak.
Since the atomic displacements of the phonon mode of interest are strictly along the crystal $c$ axis, we express the displacement vector of each carbon atom within the unit cell as
\begin{align} 
\bm{\xi}_d(\bm{Q}) =     \left(
    \begin{array}{c}
        0 \\
        0 \\
        \xi_d^z \\
        \end{array}
    \right) e^{i\varphi_d(\bm{Q})},
    \label{Eq:1_ansatz}
\end{align}
where the Cartesian $z$ axis is along the crystal $c$ axis.
The equation of the structure factor \eqref{eq:eq1} then reads:
\begin{align}
F (\bm{Q}) \propto & \sum_{d}  \hspace{.1cm} \bm{Q} \cdot\ \bm{\xi}_d (\bm{Q}) e^{i\bm{Q}\cdot\bm{r}_d} \approx\   Q_z \sum_d \hspace{.1cm} \xi_d^z e^{i(\bm{Q}_K\cdot\bm{r_d} + \varphi_d(\bm{Q}))},
\label{eq:1_1}
\end{align}
where $Q_z = \frac{2\pi}{c}L$ and $d$ runs over the four atoms within the unit cell. Here, we replace $\bm{Q}$ by $\bm{Q}_K$ in the structure factor exponents because it changes with $\bm{Q}$ only slowly and $\Delta\bm{Q} \ll\ \bm{Q}_K$. Moreover, our {\it ab initio} results demonstrate that the  magnitude of the displacements [Fig.~\ref{SFig_eigenvectors}(b)] remain almost constant in the vicinity of the K point for every $\theta$ (see the main text for the definition of the $\theta$ angle), and therefore we can safely approximate the amplitude as:
\begin{align}\label{eq:1_const_amp}
   {\xi}^z_d  :=  {\xi}^z \hspace{.5cm} \forall\ d, 
\end{align}
which further simplifies Eq.~\eqref{eq:1_1} to:
\begin{align}\label{eq:1_2}
F (\bm{Q}) \propto Q_z \xi^z \sum_d \hspace{.1cm} e^{i(\bm{Q}_K\cdot\bm{r_d} + \varphi_d(\bm{Q}))}.
\end{align}

\textit{The case of graphene.}
For simplicity, we first analyze the two-dimensional case of graphene (two atoms per unit cell) and later show that the intensity distribution remains intact when we generalize for the three-dimensional structure of graphite with four carbon ions per unit cell. The atomic coordinates in graphene are $\bm{r}_1 = (0, 0, 0)$ and $\bm{r}_2 = (\frac{1}{3}, -\frac{1}{3}, 0)$. 
%The atomic positions in graphene are $\bm{r}_1 = (0, 0, 0)$ and $\bm{r}_2 = (\frac{1}{3}, \frac{2}{3}, 0)$. 
We can recast Eq.~\eqref{eq:1_2} as:
\begin{align}
&F (\bm{Q}) \propto e^{i\varphi_1(\bm{Q})} (1 + e^{i[\bm{Q}_K\cdot\bm{r}_2+\Delta\varphi(\bm{Q})]}),
\label{eq:1_3}
\end{align}
where $\Delta \varphi (\bm{Q}) = \varphi_2(\bm{Q}) - \varphi_1(\bm{Q})$, and the constant prefactor ${Q_z} {\xi}^z$ was omitted. Because we work at arbitrary moment in time, we have freedom in choice of $\varphi_1(\bm{Q})$, which we set to 0 for every $\bm{Q}$ point without loss of generality, keeping track of the relative phase change, $\Delta \varphi (\bm{Q})$.
%To further simplify this equation without loss of generality, we can choose $\varphi_1(\bm{Q}) = 0$ for every $\bm{Q}$ point, keeping track of the relative phase change $\Delta \varphi (\bm{Q})$. 
Hence, we finally obtain an expression for the structure factor as:
\begin{align}
&F (\bm{Q}) \propto 1 + e^{i[\bm{Q}_K\cdot\bm{r}_2+\Delta\varphi(\bm{Q})]}.
\label{eq:F(Q)}
\end{align}

The IXS intensity is proportional to $|F(\bm{Q})|^2$ [Eq.~(1) from the main text]. Its calculation yields:
\begin{align}
    &|F(\bm{Q})|^2 \propto  \Re(1 + e^{i[\bm{Q}_K\cdot\bm{r}_2+\Delta\varphi(\bm{Q})]})^2 + \\ \nonumber
    &\hspace{4cm}\Im(1+ e^{i[\bm{Q}_K\cdot\bm{r}_2+\Delta\varphi(\bm{Q})]})^2; \\ \nonumber
    &|F(\bm{Q})|^2 \propto 2 
    [1 + \mathrm{cos}(\bm{Q}_K\cdot\bm{r}_2+\Delta\varphi(\bm{Q}))],
    \label{Eq:graphene}
\end{align}
which completes the derivation and show that the IXS intensity is directly governed by the phase difference $\Delta\varphi(\bm{Q})$.
Note that the phase prefactor $\bm{Q}_k\cdot\bm{r}_2 = 0$ or $\pm \frac{2\pi}{3}$ for different K points in the BZ, [$(\pm\frac{1}{3}, \pm\frac{1}{3}, 0)$ or $(\pm\frac{1}{3}, \mp\frac{2}{3}, 0)$] and accounts for the mutual orientation of the considered reciprocal-space $\bm{Q}$ point with respect to the chosen coordinate system of the unit-cell vectors in real space.

\textit{Generalization to the case of graphite.}
In the case of graphite, there are four atoms per unit cell: 
\begin{align}
 &\bm{r}_1 = (0, 0, \frac{1}{4}), &\bm{r}_2 = (\frac{1}{3}, \frac{2}{3}, \frac{1}{4}), \\ \nonumber
 &\bm{r}_3 = (0, 0, \frac{3}{4}), &\bm{r}_4 = (\frac{2}{3}, \frac{1}{3}, \frac{3}{4}).
\end{align}
Our {\it ab initio} calculations show that for the YO and BA modes the adjacent layers along the $c$ axis exhibit the anti-phase vibrations in the vicinity of K point. Taking this into account, we can express the relative phases as follow:
\begin{align} \label{Eq:phases}
&\varphi_3 = \pi + \varphi_1,         \\ \nonumber
&\varphi_4 = \pi + \varphi_2.      
\end{align}
Now we substitute Eqs.~\eqref{Eq:phases} to Eq.~\eqref{eq:1_2}:
\begin{align}\label{eq:yoba}
&F (\bm{Q}) \propto e^{i\varphi_1(\bm{Q})} 
(e^{i\bm{Q}_K\cdot\bm{r}_1} + e^{i[\bm{Q}_K\cdot\bm{r}_2
+\Delta\varphi(\bm{Q})]} + \\ \nonumber
&e^{i\bm{Q}_K\cdot\bm{r}_3} + 
e^{i[\bm{Q}_K\cdot\bm{r}_4+\Delta\varphi(\bm{Q})]}),
\end{align}
where $\Delta \varphi (\bm{Q}) = \varphi_2(\bm{Q}) - \varphi_1(\bm{Q})$. To reduce this equation to $1+ \rm{cos}(\Delta\varphi(\bm{Q}))$ form, we again choose $\varphi_1 = 0$, and specify $\bm{Q}_K = (\frac{1}{3}, -\frac{2}{3}, L)$ for $L$ odd, which corresponds to the given choice of the atomic coordinates in the unit cell.
Substituting $\bm{Q}_K$ to Eq.~\eqref{eq:yoba} we can calculate :
\begin{align} \label{eq:final}
|F(\bm{Q})|^2 \propto|e^{i\frac{\pi}{2}} + e^{i\frac{3\pi}{2}} + e^{i(\frac{\pi}{2}+\Delta\varphi(\bm{Q}))} + \\ \nonumber
e^{i(\frac{3\pi}{2}+\Delta\varphi(\bm{Q}))}|^2 =
2[1+ \mathrm{cos}(\Delta\varphi(\bm{Q}))].
\end{align}
As can be seen, the obtained result is equivalent to the one in Eq.~\eqref{Eq:graphene}. Note that to simplify calculations here we specified K $= (\frac{1}{3}, -\frac{2}{3}, 0)$, but one can easily find that for the other K points in the BZ the same dependence holds, but with the respective relative phase shift of 60$^{\circ}$.

\section{Linear dispersion}

\begin{figure}[t]
\center{\includegraphics[width=1\linewidth]{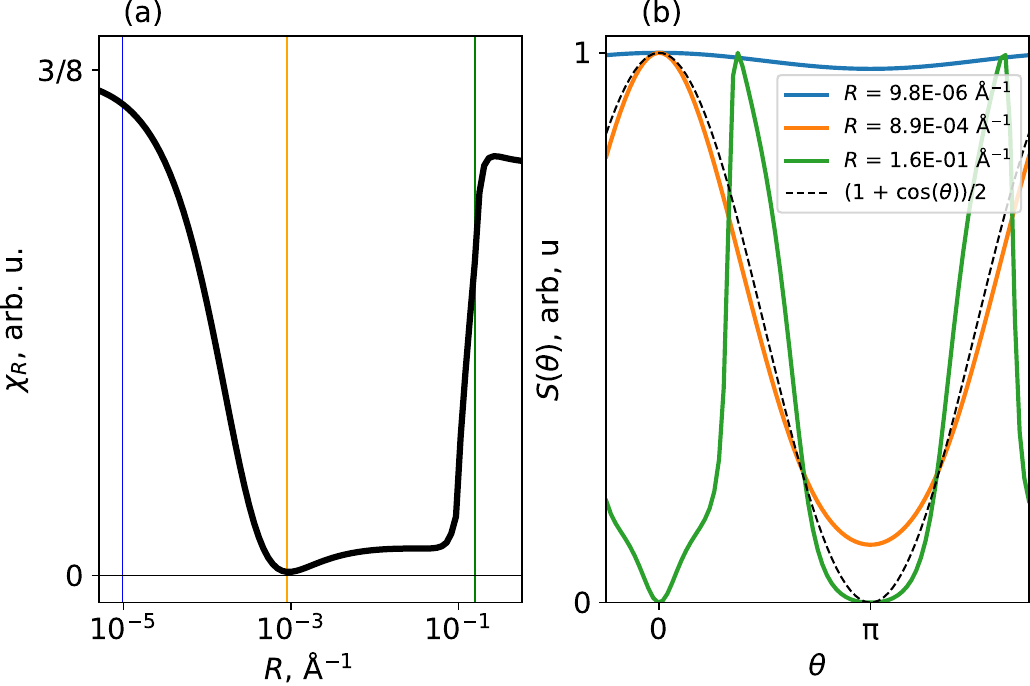}}\vspace{-12pt}
  \caption{~The dynamical structure factor of the YO mode in the vicinity of K point. (a) The $\chi$ difference as a function of radius. (b) The normalized angular dependence of the structure factor of the YO mode at a constant distance from the K point. The black dashed curve is the approximate analytical equation $(1 + \mathrm{cos}(\theta))/2$. The solid lines are the exact numerical results calculated at circles of different radii. The selected radii are also shown in (a).
  }
  \label{Schi} \vspace{-12pt}
\end{figure}

%For every $L$ we consider three pairs of the K points, $\bm{Q}_{K1} = (\pm\frac{1}{3}, \pm\frac{1}{3}, L); \bm{Q}_{K2} = (\pm\frac{2}{3}, \mp\frac{1}{3}, L); \bm{Q}_{K3} = (\mp\frac{1}{3}, \pm\frac{2}{3}, L)$.

%By substituting $\bm{r}_2$ and $\bm{Q}_{K1}$--$\bm{Q}_{K3}$ to Eq.~\eqref{Eq:nx} we find
%\begin{align}
 %   |F(\bm{Q})|^2 &\propto\ 1 + \mathrm{cos}[\Delta\varphi(\bm{Q}) \pm\ \frac{2\pi}{3}] \hspace{.5cm} 
  %  &\mathrm{for }~\bm{Q} \approx \bm{Q}_{K1}; \nonumber \\
   % |F(\bm{Q})|^2 &\propto\ 1 + \mathrm{cos}[\Delta\varphi(\bm{Q}) ]\hspace{.5cm} 
    %&\mathrm{for }~\bm{Q} \approx \bm{Q}_{K2}; \nonumber \\
    %|F(\bm{Q})|^2 &\propto\ 1 + \mathrm{cos}[\Delta\varphi(\bm{Q}) \pm\ \frac{2\pi}%{3}]\hspace{.5cm} 
   % &\mathrm{for }~\bm{Q} \approx \bm{Q}_{K3}, \nonumber \\
%\end{align}
%which complete the derivation and show that the IXS intensity is directly controlled by phase difference $\Delta\varphi(\bm{Q})$.

In the main text we discussed the dispersion of the YO and BA modes that form the pseudo-Dirac point in vicinity of the K point [see Figs.~1(c)--1(e) of the main text]. The true Dirac point in graphene is characterized by a linear crossing of the mode at the very K point, the linear slope of the Dirac phonons is then a good approximation in the vicinity of the K point and becomes less and less of a good approximation as one considers the states at momenta further away from it.

Clearly, the YO-BA modes in graphite around K posses three different regimes with respect to the associated intensity winding. First, the tiny hybridization gap, originating in the symmetry breaking between the two atoms within a honeycomb plane, affects the otherwise linearly crossed YO-BA modes up to $\approx 10^{-4}$~\AA$^{-1}$. Below this very short radius, in the other words at the very K point, the mapping onto the spin-$1/2$ model is not applicable. At some momenta greater than this, the off-diagonal terms in the $2 \times 2$ Hamiltonian of the YO-BA modes become vanishingly small so the linear dispersion becomes a good approximation and the YO-BA modes are described by the Dirac-point topology~\cite{bhattacharjee2017topology, wang2017topological, bansil2016colloquium, narang2021topology}. Similarly to graphene, the YO-BA modes that form the pseudo-Dirac cone in graphite become no longer linearly-dispersed at sufficiently large momenta away from K.

\begin{figure*}[tb]
\center{\includegraphics[width=0.9\linewidth]{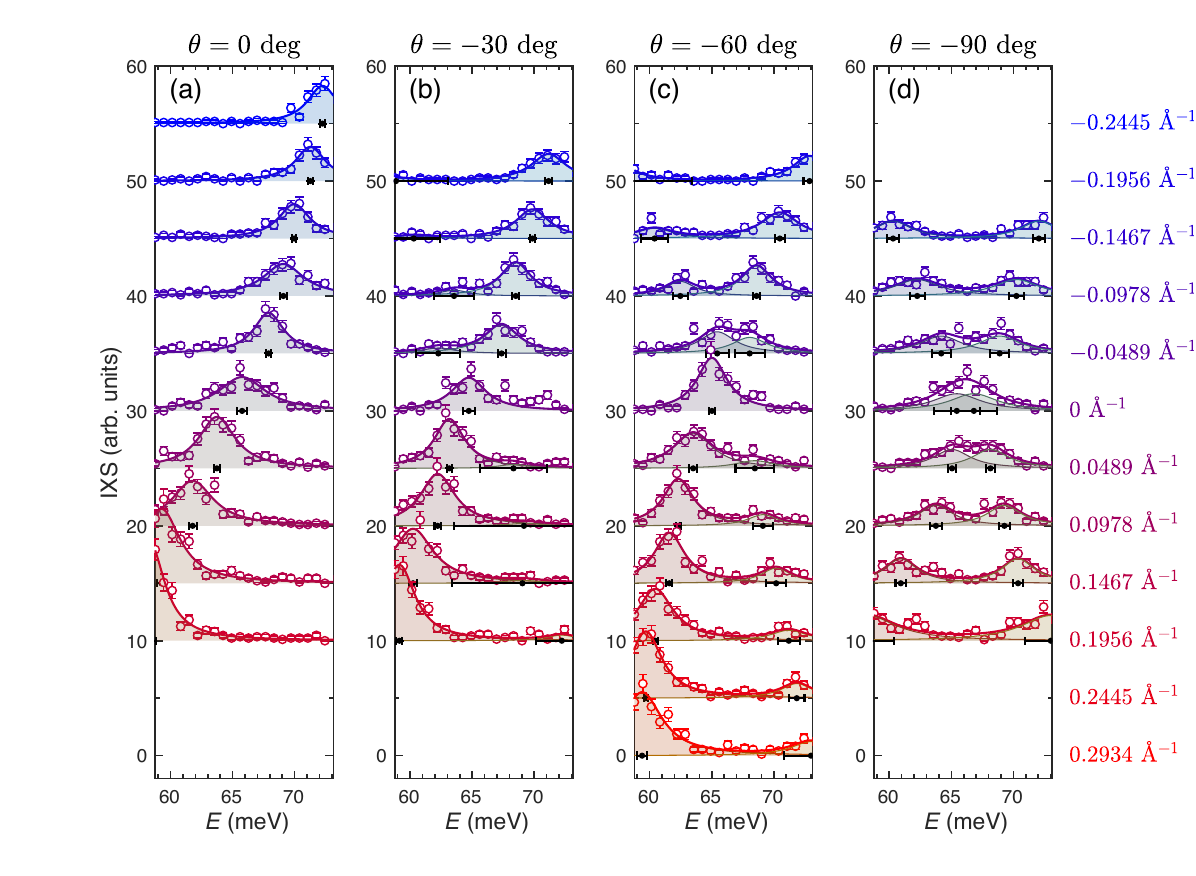}}\vspace{-12pt}
  \caption{(a)--(d) IXS spectra profiles at different momenta $\Delta \mathbf{Q}$ around the K point at $(1/3~1/3~3)$ (r.l.u.) for different $\theta$ angles within the $(hk3)$ plane (see the main text for the definition of the in-plane $\theta$ angle). The data are shown by open symbols (errorbars denote $1\sigma$ uncertainty), the solid lines are results of the fit by the Lorentzian peak profiles. The shaded areas show decomposition into the peaks of the YO and BA modes. The fitted peak positions are highlighted by the black marks.}
  \label{figS4} \vspace{-12pt}
\end{figure*} 

In order to strictly quantify the momenta range for which the topology-driven IXS intensity winding takes place, we utilize the following criteria introduced below. For this, we consider the exact variation of the IXS spectral weight of the YO mode as a function of the in-plane angle $\theta$ (defined in the main text) obtained by the numerical evaluation of the output of our {\it ab initio} lattice-dynamics simulations. The same is also applicable for the BA mode with a phase shift of $\pi$. One can numerically evaluate such exact intensity modulation on a circle in reciprocal space surrounding K and centred at K at many different radii [i.e., the distance from the K point in the $(hk0)$ plane]. The resulting intensity modulations can then be compared to the analytical expression Eq.~\eqref{eq:final} derived above. The deviation of the exact intensity modulation from the model harmonic intensity winding will therefore be in essence a measure of the mode nonlinearity.

To quantify such deviation, we consider:
%The linearity of dispersion as an attribute of Dirac point, can be quantified by calculations of the structural factor of the YO mode as a function of the angle $\phi$ at different reciprocal distances around $\bm{Q}_K = (1/3,~1/3,~3)$. A finite interlayer interaction in graphite leads to a slight splitting of modes at the $K$ point. The anticrossing results in nonlinear behaviour and deviation from the $(1 + \mathrm{cos}(\phi))$ rule in the immediate vicinity of $K$ point. As a measure of how much the spectral weight deviates from the $\mathrm{cos}({\phi})$ function, we compute the following value:

\begin{equation}
    {\chi}_R = \frac{1}{2{\pi}}\int_{0}^{2{\pi}}\biggl[\frac{(1 + \mathrm{cos}({\theta}))}{2} - S_R(\theta)\biggr]^2\,d\theta, \label{eq:chi}
\end{equation}
where $R = |\bm{Q} - \bm{Q}_K|$ represents the distance from the K point (in \AA$^{-1}$). The $R$ dependence of $\chi$ is shown in Fig.~\ref{Schi}(a). As can be seen, it exhibits three major features. The angular intensity profiles at some selected radii $R$ are represented in Fig.~\ref{Schi}(b). At $R \lesssim 10^{-5}$~\AA$^{-1}$, the entire intensity distribution tends to the limit where it is no longer dependent on the angle $S_{R\rightarrow0}(\theta) = 1$. In this limit, $\chi_{R\rightarrow0}$ approaches its maximum. The spectral weight of the YO mode at $\theta = \pi$ remains non-zero for all $R \lesssim 10^{-3}$~\AA$^{-1}$, implying that the intensity winding approximation breaks down within this momentum range. The intensity winding then takes the exact harmonic form at $R \approx 10^{-3}$~\AA$^{-1}$ where $\chi_R$ approaches zero. The intensity profile is then can be well approximated by the analytical $1 + \mathrm{cos}(\theta)$ equation for an extended range of momenta up to $R \lesssim 10^{-1}$~\AA$^{-1}$, as follows from a small deviation from the exact numerical result quantified by $\chi \ll 3/8$. For distances $R > 10^{-1}$ \AA$^{-1}$, the additional non-linear terms in the phonon dispersion begin to play a noticeable role and, as a result, introduce the higher harmonics in the intensity winding. Above this upper limit, the analytical cosine equation is no longer a good approximation of the intensity modulation of the YO-BA modes. One should also bear in mind that at a circular path in reciprocal space with such a large radius, the energy of the YO and BA modes is no longer constant, which as well results in a more complex non-harmonic behaviour of the structure factor and thus growing of the calculated $\chi$.

%At $R \lesssim 10^{-3}$~\AA$^{-1}$, the spectral weight of the YO mode at $\phi = \pi$ becomes non-zero and the entire curve tends to the limit where intensity is no longer depends on angle $S_{R\rightarrow0}(\phi) = 1$ (blue curve in Fig.~\ref{Schi}(b)). In this limit, $\chi_{R\rightarrow0}$ approaches to 3/8. For distances $R > 10^{-1}$ \AA$^{-1}$, the effect of non-linear dispersion begins to play a role (green curve in Fig.~\ref{Schi}(b)). In a circular path in reciprocal space with such a radius, the energy of the YO mode is no longer constant, which as well results in more complex behaviour of structure factor and growing of $\chi$. In the intermediate region, where $\chi$ is rather small, the $S(\phi)$ is reliably described with a $(1 + \mathrm{cos}(\phi))$ function (orange curve in Fig.~\ref{Schi}(b)).

\section{Fitting of the phonon dispersion} \label{sec:fit}

Figures~2(b,c,d,e) of the main text show the observed energy-momentum IXS spectra maps in vicinity of the K point at $\mathbf{Q} = (1/3~1/3~3)$~(r.l.u.). In order to directly compare the obtained dispersion of the YO and BA modes (which feature the quasi-Dirac point at K) with the result of our {\it ab initio} calculation, we fitted each individual intensity profile $I$ as a function of the energy $E$ with a Lorentzian peak. The Lorentzian peak function has three free parameters, which are the peak position, the amplitude of the peak, and the peak width. A sum of two Lorentzian peaks were used to fit the profiles in Figs.~\ref{figS4}(b--d), whereas only one Lorentzian peak was used for the intesity profiles in Fig.~\ref{figS4}(a). As a result of these fits, we extracted the respective peak positions and plotted them over the experimental as well as simulated momentum energy IXS spectra in Figs.~2(b)--2(i) of the main text. The errorbars of the extracted peak positions denote 95\% confidence interval obtained through the fitting procedure. Figures~\ref{figS4}(a--d) summarize the results of all the fits used to plot Figs.~2(b--i) of the main text.

\begin{figure}[tb]
\center{\includegraphics[width=0.99\linewidth]{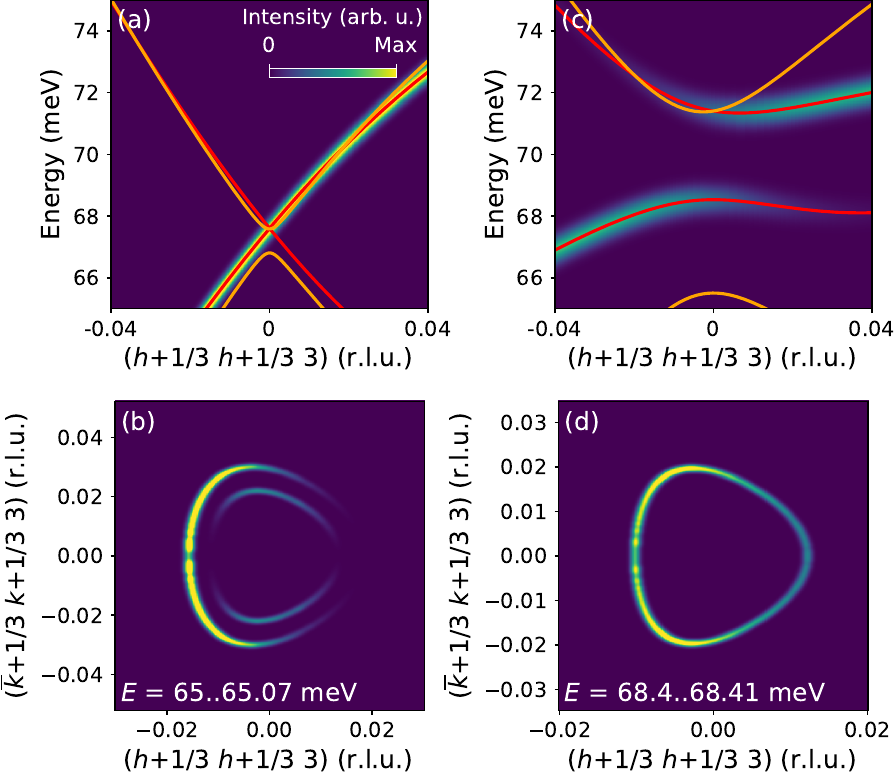}}\vspace{-5pt}
  \caption{IXS intensity slices along $(hh3)$ and constant-energy slice in graphite in absence of external pressure (a,b) and under hydrostatic pressure of 50~GPa (c,d). The maximum of the colormap intensity scale was adjusted respectively in each panel for better feature visibility. 
  }
  \label{figS5} \vspace{-10pt}
\end{figure}

\section{Winding of IXS intensity for stronger interlayer coupling} \label{sec:fit}
Figure~1 (c) of the main text demonstrates phonon dispersion of graphite, and panel (d) shows a region close to the $K$ point where YO and BA modes are. In the main text we studied winding of spectral weight in the close vicinity of $\mathbf{Q} = (1/3~1/3~3)$~(r.l.u.), where the intensity of YO-BA modes is maximized. Simulated dispersion of these modes exhibits a weak gap at the K point, which means that from experimental point of view they behave indistinguishably from the true Dirac cone. However, for a larger gap this conclusion will no longer be valid. Here, we show how the harmonic modulation of spectral weight breaks down in the case of a large gap between YO and BA modes at the $K$ point.

In graphite with weak van-der-Waals forces that couple the layers, the value of the gap is driven by the strength of those interactions and can be naturally controlled by application of external pressure. This is evident from the lattice parameter measurements under hydrostatic pressure~\cite{zhao1989x}. The measurements demonstrate that the lattice parameter $c$ strongly decreases already under moderate pressures ($P < 10$~GPa), when the in-plane parameter $a$ only barely changes. Thus, applied pressure can serve as a convenient tuning knob to study the relations between the gap size and the spectral weight distribution around the $K$ point, as well as the limits in which the dispersion of the YO and BA modes is linear.

In order to address the question of how the gap size influences the observed winding of IXS intensity in graphite, we performed \textit{ab initio} calculations within the same framework as for the results in the main text, but under simulated high hydrostatic pressure. For this, the crystal structure was let to relax, which resulted in the lattice parameter $c$ of 6.485~$\text{\AA}$ in the absence of applied pressure and a reduced value of 4.992~$\text{\AA}$ under applied pressure of 50~GPa.

In Fig.~\ref{figS5}(c) we present the phonon dispersion calculated at 50~GPa in the vicinity of the former YO-BA weak anti-crossing. As can be seen, the gap has increased substantially with respect to the original (no pressure) states. This is 50~$\mu$eV at 0~GPa (see the main text) and 2.9~meV at 50~GPa. In Figs.~\ref{figS5}(b) and ~\ref{figS5}(d) we show calculated constant-energy slices in the $(hk3)$ plane close to the $K$ point without [Fig.~\ref{figS5}(b)] and with [Fig.~\ref{figS5}(d)] applied pressure. The energy was chosen such that it lies just below the pseudo-Dirac point at ambient pressure. The constant-energy slice at pressure of 50~GPa is shown for the same energy for a direct comparison. Clearly, the zero-pressure pattern demonstrates a winding of the spectral weight, whereas the pressure-driven state with a large gap is characterized by a completely reorganized distribution of the spectral weight, which is non-zero everywhere in the plane. This shows that the pseudo-Dirac states are no longer a valid approximation for the phonon band structure in graphite when the interlayer interactions are strong and the gap between YO and BA modes becomes large.

%case the intensity is distributed \red{do some zoology!}, indicating breakdown of intensity \red{winding scenario when the gap is large compared to the resolution. }

\bibliography{bibliography}